\documentclass[letterpaper, 10 pt, conference]{ieeeconf}  

\IEEEoverridecommandlockouts                              
\usepackage{amsmath,amssymb,amsfonts}
\usepackage{bm,bbm,siunitx,hyperref,graphicx}
\usepackage{float}	
\usepackage{xcolor}

\let\labelindent\relax

\usepackage{enumitem}
\usepackage{makecell}
\usepackage{cellspace}
\usepackage{cite}

\usepackage[skip=10pt,font=small]{caption}
\usepackage{balance}   

\usepackage{soul}	

\DeclareSIUnit \ampereHour {Ah}

\title{\LARGE \bf
Flying batteries: In-flight battery switching \\ to increase multirotor flight time
}

\author{
Karan P. Jain and Mark W. Mueller
\thanks{Authors are with the High Performance Robotics Laboratory (HiPeRLab) at the Department of Mechanical Engineering, UC Berkeley, CA 94720, USA.
{\tt\small \{karanjain, mwm\}@berkeley.edu}} }
\begin{document}
 
\maketitle

\begin{abstract}
We present a novel approach to increase the flight time of a multirotor via mid-air docking and in-flight battery switching.
A main quadcopter flying using a primary battery has a docking platform attached to it.
A `flying battery' -- a small quadcopter carrying a secondary battery -- is equipped with docking legs that can mate with the main quadcopter's platform.
Connectors between the legs and the platform establish electrical contact on docking, and enable power transfer from the secondary battery to the main quadcopter.
A custom-designed circuit allows arbitrary switching between the primary battery and secondary battery.
We demonstrate the concept in a flight experiment involving repeated docking, battery switching, and undocking.
This is shown in the video attachment\footnote{The explanation and experimental validation video can be found here: \url{https://youtu.be/PpJIo4HXl_Q}}.
The experiment increases the flight time of the main quadcopter by a factor of $4.7\times$ compared to solo flight, and $2.2\times$ a theoretical limit for that given multirotor.
Importantly, this increase in flight time is not associated with a large increase in overall vehicle mass or size, leaving the main quadcopter in fundamentally the same safety class.
\end{abstract}

\section{Introduction}\label{sec:intro}
Multirotors are frequently employed in mapping, delivery, monitoring, search and rescue missions \mbox{\cite{sharma2016uav, waharte2010supporting, erdelj2017help, cacace2016control}} among many other applications owing to their ability to hover.
However, multirotors inherently have lower endurance and range as compared to fixed-wing aircraft \cite{boon2017comparison}.
There is a growing demand for higher endurance and range in multirotors with their increasing usage in the research and industrial setting.

Current literature covers innovative methods to increase the endurance of multirotors.
A hybrid aerial vehicle is presented in \cite{thamm2015songbird} which exploits the efficiency of a fixed-wing and hovering ability of a multirotor.
An online strategy for optimizing efficiency by altering flight parameters over a trajectory is presented in \cite{tagliabue2019model}.
An approach without attempting to increase efficiency is to have a recharging station for a quadcopter, demonstrated for example in \cite{junaid2017autonomous}.
As recharging is substantially slower than conventional refuelling, instead a discharged battery may be swapped with a charged one.
Battery swapping at a ground station has been shown in \cite{lee2015autonomous, toksoz2011automated, ure2014automated}.
One limitation of ground-based swapping stations is an interruption to the mission.
For example, if a quadcopter's mission is monitoring a target, then going to a ground station for battery replacement results in a mission failure.

A spare battery having the ability to \emph{fly} to the quadcopter will enable an uninterrupted mission.
This capability can be enhanced if the spare battery can be removed in-flight after discharging so that another spare battery can take its place and continue providing energy.
Moreover, this would allow a system to operate for long-distance flights without the disruption of stopping the flight, potentially a crucial feature to applications such as urban air mobility.

\begin{figure}
    \centering
    \includegraphics[width=\columnwidth]{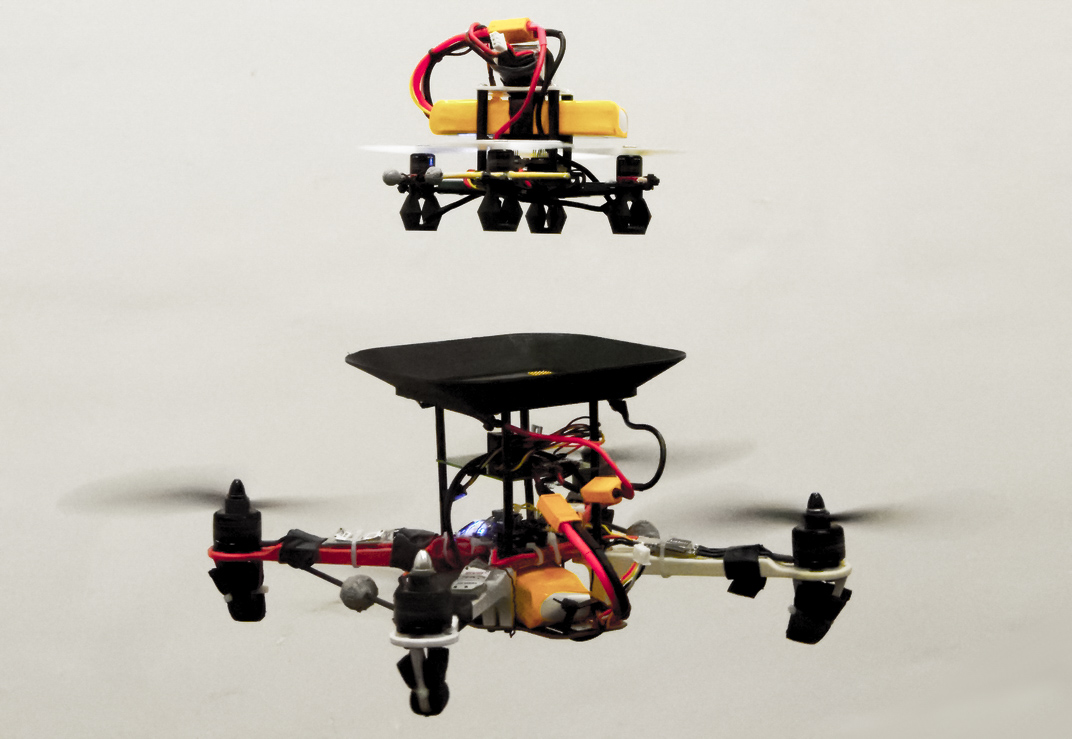}
    \caption{A flying battery (above) about to dock on the main quadcopter (below).}
    \label{fig:LQ_MQ_docking}
\end{figure}

We present the concept of a `flying battery' -- a secondary battery mounted on a small quadcopter.
While a main quadcopter performs some task mid-air using a primary battery, a flying battery can fly towards the main quadcopter and dock on it.
The main quadcopter can then switch its power source to the secondary battery.
Once the flying battery is depleted, it can undock, and another fully charged flying battery can dock in its place.
This process can be repeated until the primary battery, which is only used from the time when one flying battery undocks until another one docks back, is depleted.
This increases the total flight time and is achieved while the main quadcopter is airborne, so there is no interruption to the mission.
Fig. \ref{fig:LQ_MQ_docking} shows a flying battery approaching the main quadcopter to dock on it.

The paper is organized as follows.
Section \ref{sec:motivation} presents a fundamental flight time vs. battery mass analysis to motivate the usage of our proposed system.
Section \ref{sec:design} explains the hardware design of our system.
Section \ref{sec:docking} covers the method of docking a flying battery on the main quadcopter and undocking it.
Section \ref{sec:demo} demonstrates how our design increases the flight time of the main quadcopter.

\newcommand{\flightTime}{T_\mathrm{flight}}
\newcommand{\batteryEnergy}{E_\mathrm{batt}}
\newcommand{\electricPowerDraw}{p_\mathrm{elec}}
\newcommand{\baseMass}{m_0}
\newcommand{\energyDensity}{\gamma}
\newcommand{\powerMassRatio}{k_p}

\section{Motivation}\label{sec:motivation}
In this section, we present an analysis of a fundamental limitation of hovering battery-powered multirotors.
Specifically, we show that the achievable flight time only increases up to a certain point, as more battery is used on a vehicle. 

We model the aerodynamic power consumption $p_i$ of an individual propeller $i$ to be related to its thrust $f_{i}$ as 
\begin{align}
	p_i \propto f_{i}^{\frac{3}{2}}\label{eqPowerAndThrust}
\end{align}
This can be derived from actuator disk theory \cite{mccormick1995aerodynamics}, or from mechanical analysis of hub torque and rotational speed \cite{holda2018tilting}.

Assuming constant energy density $\energyDensity$, the total available energy $\batteryEnergy$ in a battery will be $\energyDensity$ times the battery's mass.

Let $\baseMass$ be the mass of all components of the multirotor, excluding the battery, and let $\phi$ be the fraction of the \emph{total} vehicle mass that is the battery mass, so that the total vehicle mass is $\frac{1}{1-\phi}\baseMass$ and the battery mass is $\frac{\phi}{1-\phi}\baseMass$.

For a hovering multirotor, the individual propeller thrusts scale proportionally with the vehicle's total mass.
The electric power draw $\electricPowerDraw$ can be written as,
\begin{align}
\electricPowerDraw = \powerMassRatio \left(\frac{1}{1-\phi}\baseMass\right)^{3/2}
\end{align}
where $\powerMassRatio$ is determined by propeller design, gravitational acceleration, and powertrain efficiency.
We assume these parameters are constant for a given vehicle.

The available flight time $\flightTime$ can then be related to the vehicle mass $\baseMass$, and battery mass fraction:
\begin{align}
	 \flightTime = \frac{\batteryEnergy}{\electricPowerDraw}
	 = \frac{\energyDensity\frac{\phi}{1-\phi} \baseMass}{\powerMassRatio \left(\frac{1}{1-\phi}\baseMass\right)^{3/2}}
	 \propto \frac{\phi\sqrt{1-\phi}}{\sqrt{\baseMass}}
	 \label{eqFlightTimeFromBattMass}
\end{align}

\begin{figure}
    \centering
    \includegraphics[width=\columnwidth]{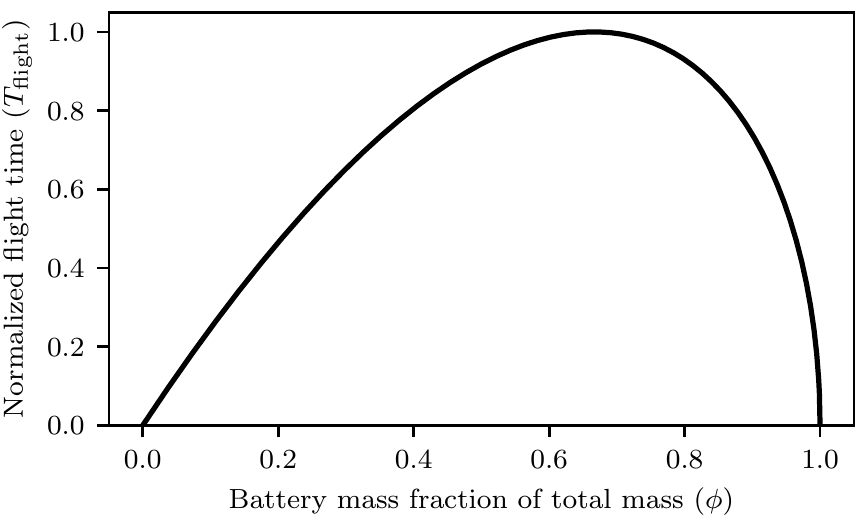}
    \caption{Effect of battery mass on normalized hovering flight time. After the peak at $\phi=\frac{2}{3}$, a larger battery reduces flight time.}
    \label{fig:battWt}
\end{figure}

This relationship is plotted in Fig.~\ref{fig:battWt}, showing that vehicles with relatively small batteries expect to see a strong improvement in total flight time with increasing battery mass, until a peak where the battery takes up two-thirds of the vehicle's mass.
We note that the location of this peak is independent of $\energyDensity$ and $\powerMassRatio$.
This large fraction makes structural design difficult and may lead to potential safety concerns. 
This analysis motivates our proposed system -- by creating a system that enables the multirotor to ``shed'' a discharged battery, and replace it with a fully charged battery, the vehicle is able to exceed the flight-time limitation imposed by \eqref{eqFlightTimeFromBattMass}.

\section{Design}\label{sec:design}

In this section, we explain the design of the mid-air docking mechanism, the battery switching circuit, and the quadcopters used in our experiments.

\subsection{Docking mechanism}\label{sec:dockMech}
\begin{figure}[hb]
	\centering
	\includegraphics[width=\columnwidth]{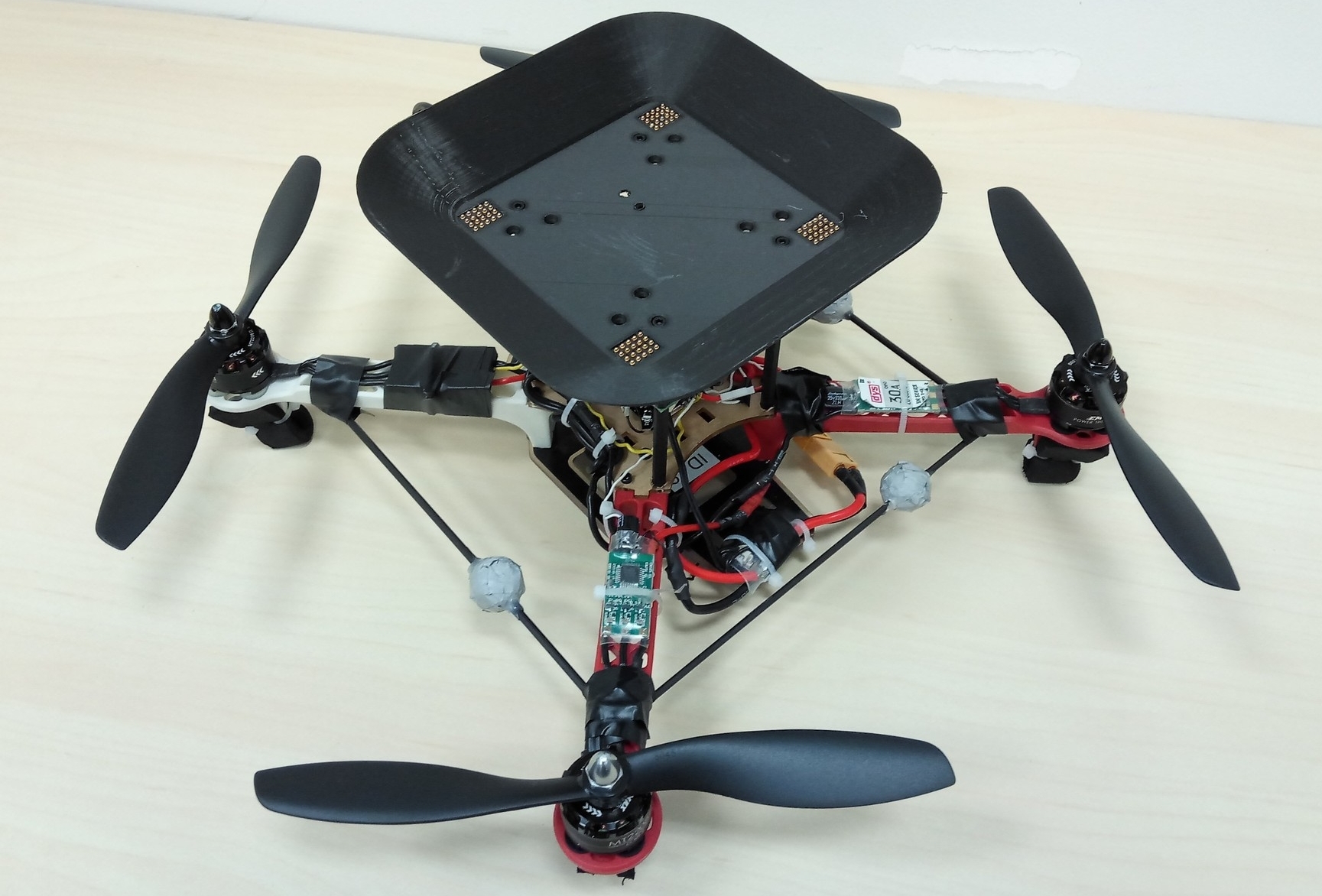}\\
	\vspace{1mm}
	\includegraphics[width=\columnwidth]{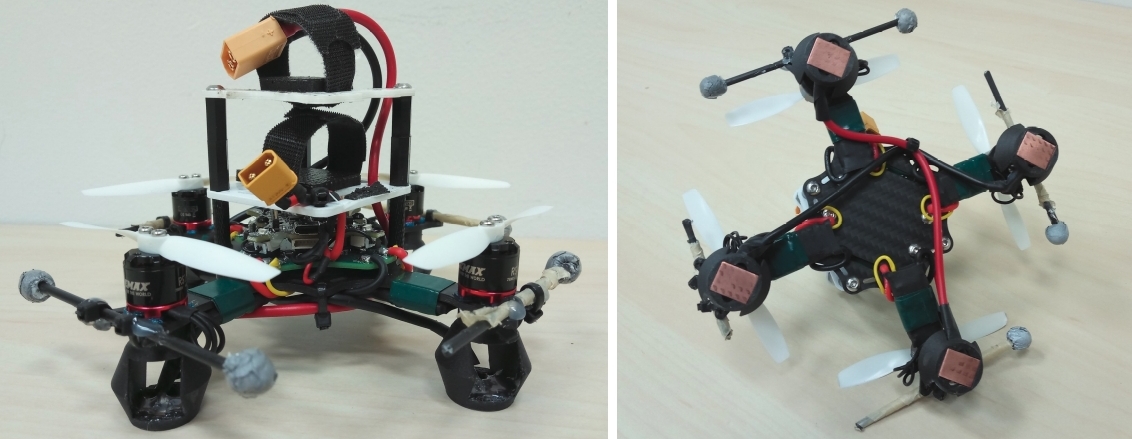}
	\caption{\emph{Top:} Main quadcopter with the docking platform and spring loaded connectors. \emph{Bottom:} Flying battery with the docking legs and copper plate connectors.}
	\label{fig:LQandMQ}
\end{figure}

Mid-air docking of multirotors has been performed using a variety of mechanisms.
Robotic hands, a winch, and a rod were used by \cite{miyazaki2018airborne} to dock vehicles vertically aligned.
Lateral docking using magnets has been demonstrated in \cite{saldana2018modquad}.
Our proposed solution uses a mechanical guide structure in the form of a landing platform on the main quadcopter and landing legs on the flying battery as shown in Fig. \ref{fig:LQandMQ}. This enables a fast docking procedure along with an easy undocking process.
This design achieves the following objectives:
\begin{itemize}
    \item \textit{No active components:} The mechanism does not consume power, and uses the weight of the flying battery for docking.
    This makes it light-weight and leads to a simple undocking process -- regular take-off.
    \item \textit{Docks vertically aligned:} The flying battery does not produce any thrust when docked.
    Aligning the center of mass of the docked configuration along the thrust direction of the main quadcopter prevents unbalanced thrusts and additional power consumption.
    \item \textit{Precision landing:} We require a secure electrical contact after docking to power the main quadcopter from the flying battery.
    This necessitates the electrical connectors to be well aligned.
\end{itemize}

The docking platform and the legs include electrical connectors which can allow the transfer of power from a flying battery to the main quadcopter.
The docking mechanism allows some lateral play between the vehicles to facilitate smooth docking and undocking, but this play is limited sufficiently to ensure that the electrical connections are not broken due to vibrations and dynamic motions.

\begin{figure}
	\centering
	\includegraphics[width=0.5\columnwidth]{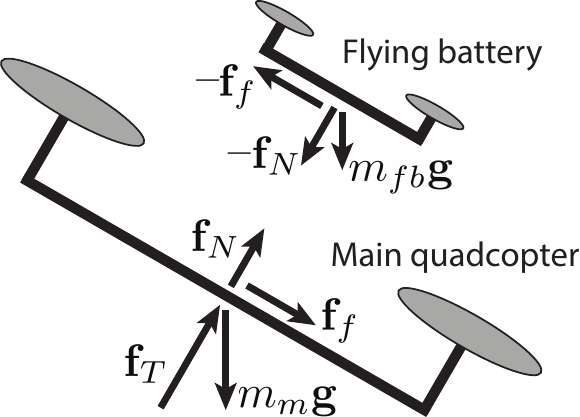}
	\caption{Free body diagram of the two vehicles when docked}
	\label{fig:FBDmaneuver}
\end{figure}

The increased mass and moment of inertia of the main quadcopter in the docked configuration reduces its agility.
However, it can still perform moderately agile motion while maintaining electrical contact, even without an active fixing mechanism.
This can be analyzed by computing the required friction for zero relative acceleration between the two vehicles when docked.
We assume, for simplicity, that rotary motion is negligible, and that the only forces acting on the main quadcopter are its weight, total thrust from the propellers $\mathbf{f}_T$, and the normal $\mathbf{f}_N$ and friction $\mathbf{f}_f$ force at the docking mechanism, as is shown in Fig.~{\ref{fig:FBDmaneuver}}.
The flying battery is acted upon by its weight, and the normal and friction forces from the docking mechanism (equal and opposite to those acting on the main quadcopter).
The vector of gravity is $\mathbf{g}$.
Letting $m_m$ and $m_{fb}$ respectively be the mass of the main quadcopter and the flying battery, we can compute the requirement for zero relative acceleration by applying Newton's law, where $\mathbf{a}_m$ and $\mathbf{a}_{fb}$ are the acceleration of the main quadcopter and flying battery, respectively:

\begin{align}
	\mathbf{0} &= \mathbf{a}_m - \mathbf{a}_{fb} 
	\\&= \frac{m_m \mathbf{g} + \mathbf{f}_T + \mathbf{f}_N + \mathbf{f}_f}{m_m}
	- \frac{m_{fb} \mathbf{g} - \mathbf{f}_N - \mathbf{f}_f}{m_{fb}}
	\\&=\frac{\mathbf{f}_T}{m_m} + \mathbf{f}_N \frac{m_m+m_{fb}}{m_m m_{fb}}  + \mathbf{f}_f \frac{m_m+m_{fb}}{m_m m_{fb}} 
\end{align}

Because the body-fixed thrust direction is normal to the docking platform, the forces $\mathbf{f}_T$ and $\mathbf{f}_N$ are parallel, and perpendicular to $\mathbf{f}_f$.
As long as the thrust produced by the main quadcopter is positive (as would be true under almost all flight conditions), the normal force at the platform will act into the platform (ensuring electrical contact), and no friction force is required for the vehicles to remain docked.
In reality, there will be additional forces acting on the system, such as aerodynamic drag during translation.
In such cases, the required friction force is non-zero.
Large angular velocities or angular accelerations may lead to loss of normal force.
However, placing the docking platform close to the main quadcopter's center of mass makes this unlikely to occur.

\subsection{Battery switching circuit}\label{sec:batterySwitch}
A crucial feature of our design is seamless switching between the primary battery and secondary battery.
Since our system is flying, we cannot afford to cut the power supply during this switch.
The two batteries need to be connected in parallel for some time to achieve this, and this is only safe within a voltage difference of \SI{0.2}{\volt} per cell for lithium polymer (LiPo) batteries.
This would often not be the case in our application because we intend to utilize the secondary battery from a fully charged state (\SI{4.2}{\volt} per cell) to a completely discharged state (\SI{3.0}{\volt} per cell).
We solve this by connecting diodes in series with each of the batteries to avoid reverse currents.
We utilize smart bypass diodes because they have a much lower voltage drop than conventional P-N junction or Schottky diodes.
At our operating current of about \SI{18}{\ampere}, the voltage drop is less than \SI{0.1}{\volt}.

\begin{figure}
	\centering
	\includegraphics[width=\columnwidth]{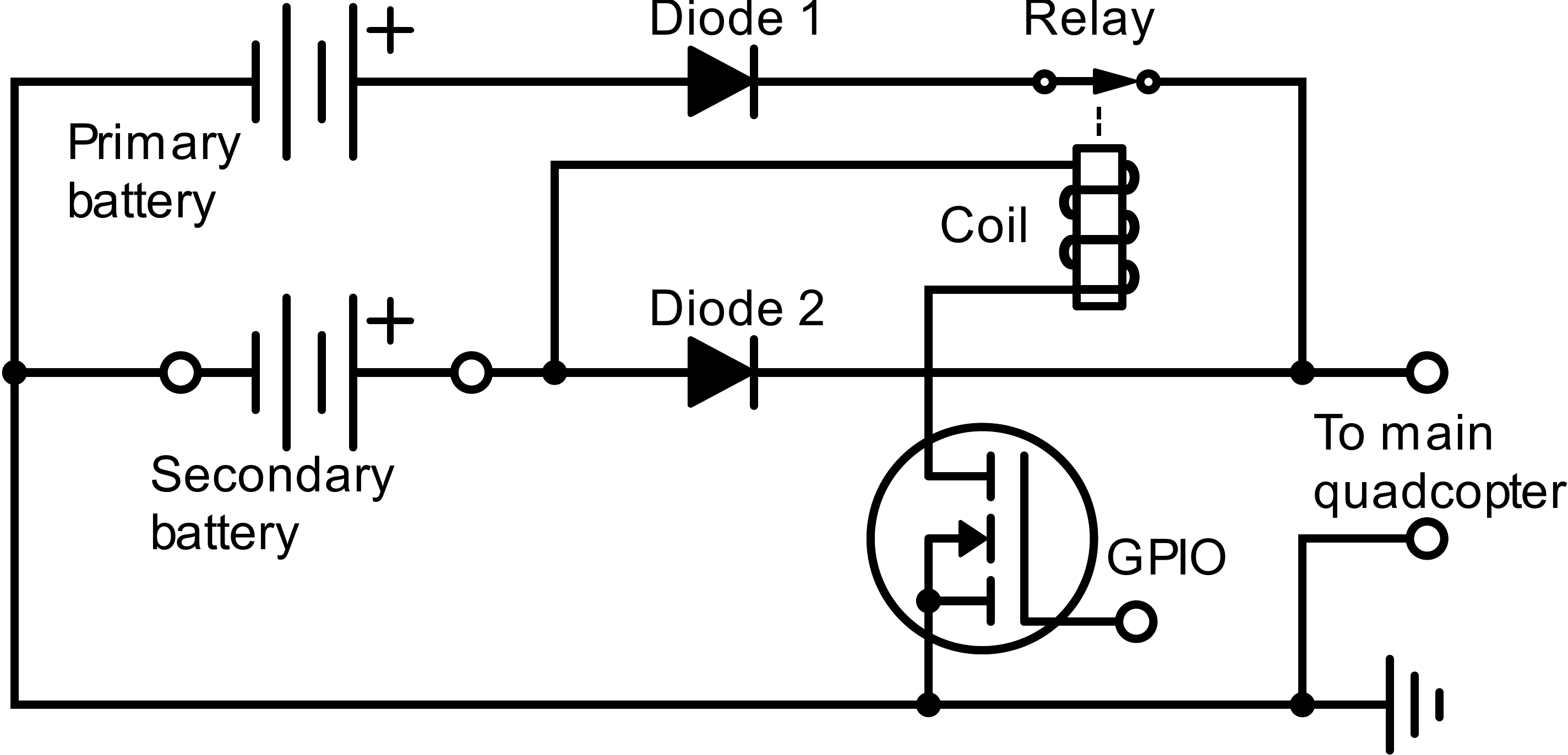}
	\caption{Schematic of the battery switching circuit.}
	\label{fig:battSwitcher}
\end{figure}

A normally closed relay is connected in series with the primary battery.
By opening the switch, we can draw power from the secondary battery even when it is at a lower voltage than the primary battery.
The relay coil is connected across the secondary battery input leads in series with a MOSFET.
This ensures that the switch does not open without a secondary battery, and allows us to use a GPIO pin on the flight controller to control the switch.
Fig. \ref{fig:battSwitcher} shows a schematic diagram of the battery switching circuit.

\subsection{Vehicle design}\label{sec:vehDesign}

\subsubsection{Main quadcopter}\label{sec:LQ}
The main quadcopter is designed to have enough payload capacity for carrying useful sensors such as surveillance cameras, or environmental sensors. 
The battery switching circuit and docking platform are stacked on top of the main quadcopter.
Spring-loaded connectors are mounted on the docking platform to serve as input leads to the quadcopter from the secondary battery.
The primary battery is a 3S \SI{2.2}{\ampereHour} LiPo battery, which weighs \SI{190}{\gram}.

\subsubsection{Flying battery}\label{sec:MQ}
The small quadcopter is designed to have sufficient payload capacity to carry a secondary battery for the main quadcopter.
The docking legs for the small quadcopter are designed to minimize blockage of the propeller airflow to minimally affect the payload capacity.
Copper plates of dimensions similar to the spring-loaded connectors are installed on the legs to serve as the secondary battery output leads.
The small quadcopter is powered using a 2S \SI{0.8}{\ampereHour} LiPo battery weighing \SI{45}{\gram}.
The secondary battery is a 3S \SI{1.5}{\ampereHour} LiPo battery, which weighs \SI{135}{\gram}.

Table \ref{table:vehicleSpec} summarizes the specifications of the two vehicles.

\begin{table}
	\begin{centering}
		\caption{Specifications of quadcopters used in experiments}
		\label{table:vehicleSpec}
		\begin{tabular}{|c|c|c|}
			\hline
			\thead{Parameter}			& \thead{Main quadcopter}	& \thead{Flying battery}\\
			\hline Propeller diameter	& \SI{203}{\mm}			& \SI{76}{\mm}
			\tabularnewline
			\hline Arm length			& \SI{165}{\mm}			& \SI{58}{\mm}
			\tabularnewline
			\hline Mass					& \SI{820}{\gram}		& \SI{320}{\gram}
			\tabularnewline
			\hline Maximum thrust		& \SI{27}{\newton}		& \SI{8}{\newton}
			\tabularnewline
			\hline
		\end{tabular}
	\par\end{centering}
\end{table}


\section{Docking Maneuver}\label{sec:docking}
This section covers the considerations involved in docking and undocking the flying battery and the main quadcopter.

\subsection{Aerodynamic disturbance rejection}\label{sec:aerodRej}
A critical consideration for docking two quadcopters mid-air is the mutual aerodynamic interference caused by the airflow of the two vehicles, especially during vertical docking because one quadcopter is directly in the downwash of another.
An analysis of rotorcraft downwash is presented in \cite{yeo2015empirical, yoon2016computational}.
Detailed characterization and analysis of aerodynamic forces and torques between two quadcopters is shown in \cite{jain2019modeling}.
We will use the following key results from \cite{jain2019modeling}:
\begin{enumerate}
  \item The effect of mutual aerodynamic disturbances is primarily seen on the quadcopter that flies below.
  The quadcopter that flies above is negligibly affected.
  \item The predominant component of the aerodynamic forces is along the direction of the downwash.
  Forces perpendicular to the direction of the downwash can be ignored.
  \item The aerodynamic torques disturb the bottom quadcopter in a way that tends to vertically align it with the top quadcopter.
  This is advantageous in our maneuver.
  Hence, we do not attempt to reject the torques.
\end{enumerate}

We chose to fly the flying battery above the main quadcopter owing to result (1).
The main quadcopter has sufficient thrust capacity to reject the disturbances caused by the flying battery's airflow.
Based on results (2) and (3), the only disturbance that we correct for is the vertical force.
This is done by applying a feedforward thrust based on the relative location of the two quadcopters.

The feedforward thrust map is created by flying the two quadcopters at various relative separations.
A PID controller is used for position control which outputs a desired total thrust force.
The map is created from previous runs' controller integral actions and tested on subsequent runs until satisfactory disturbance rejection is achieved.

\subsection{Docking trajectory}\label{sec:dockTraj}
The key requirement of this project is that the main quadcopter should not have to move substantially from its place during a long-term operation. 
Therefore, the docking trajectory involves minimal motion of the main quadcopter.
Result (3) in Section \ref{sec:aerodRej} mentions that aerodynamic torques tend to vertically align the two quadcopters.
Based on this, we start by commanding the flying battery to go \SI{30}{\centi\meter} vertically above the docking platform on the main quadcopter.
It is then commanded to descend towards the docking platform.
In this phase, any misalignments and tracking errors are corrected by the aerodynamic torque on the main quadcopter.

Once the flying battery's center is within \SI{2.0}{\centi\meter} of the docking platform's center in the horizontal plane, and the bottom surface of its legs is within \SI{5.0}{\centi\meter} of the platform's surface, it is commanded to free fall.
The docking platform is designed to precisely align the flying battery's connectors with those on the main quadcopter once they are within a \SI{2.0}{\centi\meter} radius in the horizontal plane.
The drop height of \SI{5.0}{\centi\meter} was chosen to have sufficient impact for the flying battery to slide in and align correctly, and also avoid rebounding which might cause misalignment.
Fig. \ref{fig:LQ_MQ_docking} shows a picture of the flying battery about to dock on the main quadcopter.
Starting from takeoff, it takes \SI{20}{}-\SI{25}{\second} for the flying battery to dock on the main quadcopter.
These distances and times were experimentally determined to give satisfactory results.

For undocking, we command the flying battery to takeoff from the docking platform and go straight up to a position \SI{30}{\centi\meter} above the platform.
After this, the flying battery lands and another one is free to dock on the main quadcopter.
Including landing, this maneuver takes approximately \SI{8}{\second}.

\section{Experimental Validation}\label{sec:demo}

\begin{figure}[b]
	\centering
	\includegraphics[width=\columnwidth]{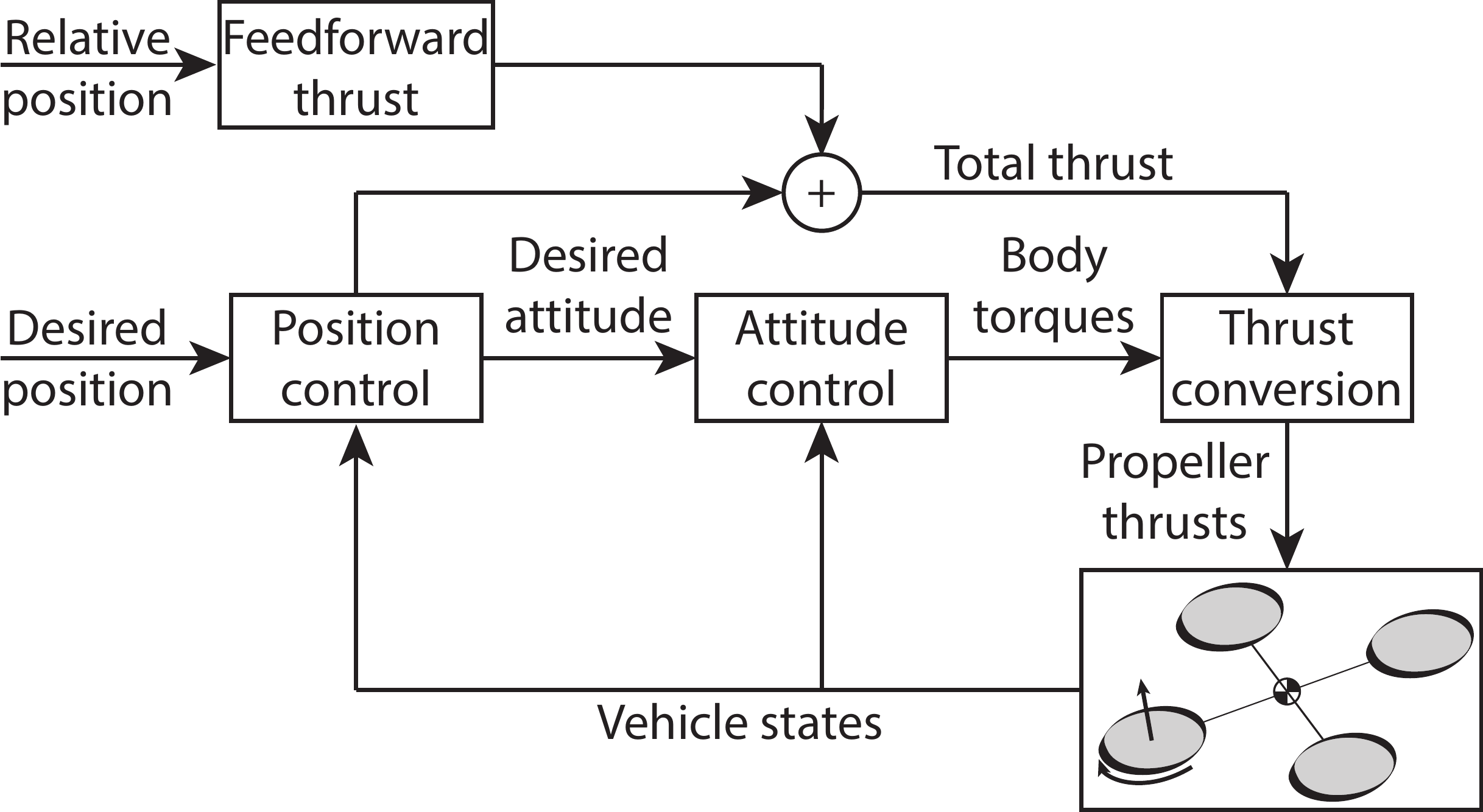}
	\caption{Block diagram of the quadcopter controller. The flying battery does not have the feedforward thrust component.}
	\label{fig:ctrler}
\end{figure}

\begin{figure*}
	\centering
	\includegraphics[width=\textwidth]{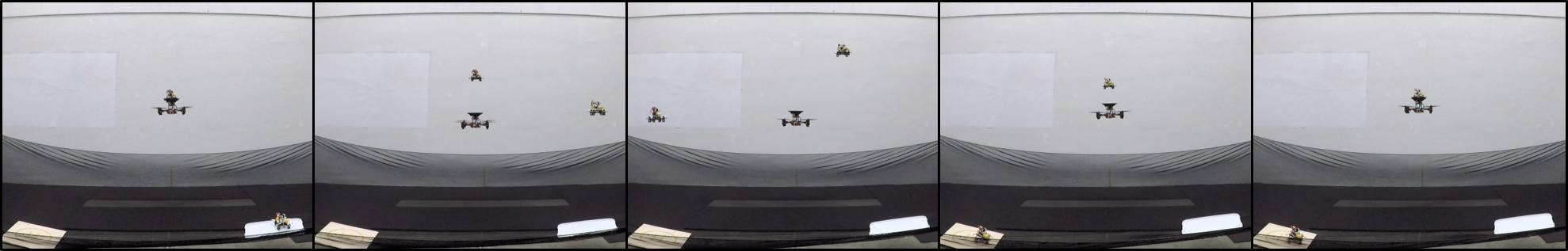}
	\caption{Steps (4)-(6) of the demonstration. From left to right,
		(a) main quadcopter hovers with a flying battery docked on it,
		(b) the first flying battery is depleted, so it undocks and another fully charged flying battery takes off,
		(c) the second flying battery moves towards the main quadcopter to dock and first flying battery begins landing,
		(d) second flying battery descends to dock on the main quadcopter,
		(e) second flying battery is docked on the main quadcopter (which continues to hover) and first flying battery has landed - we now manually replace the discharged flying battery with a fully charged one.}
	\label{fig:dockTwoMQReel}
\end{figure*}

\begin{figure*}
	\centering
	\includegraphics[width=\textwidth]{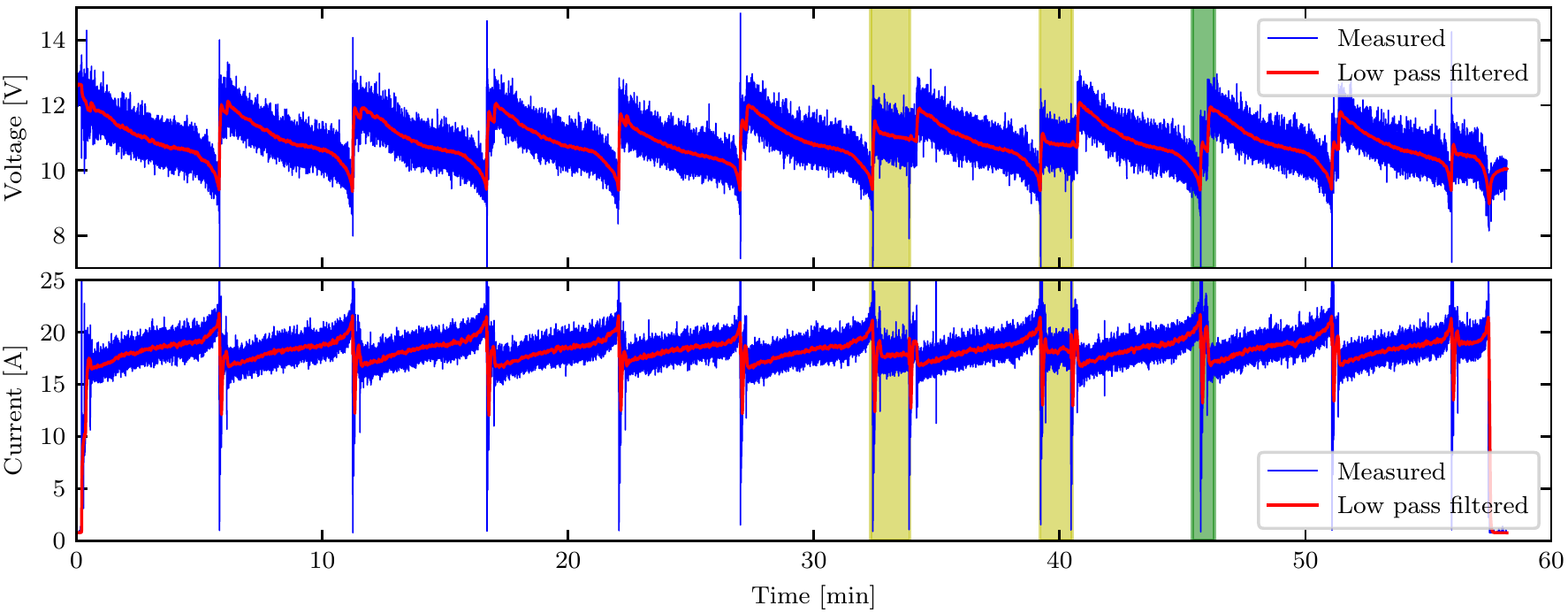}
	\caption{Input voltage and current vs. time of the main quadcopter for the demonstration.
		Low pass filtered data is also plotted for ease of visualization.
		The maximum current value of \SI{55}{\ampere} is cut off in the plot.
		Regions highlighted in yellow show parts of the experiment where the flying battery did not make electrical contact after docking.
		Fig. \ref{fig:battVIZoomed} shows a zoomed-in version of the green highlighted region.}
	\label{fig:battVI}
\end{figure*}

We validate the use of our design by conducting an experiment involving repeated docking, battery switching, and undocking so that the main quadcopter's use of the primary battery is only limited to the undocking and docking phases.
Whenever a flying battery is docked correctly, we use the secondary battery to power the main quadcopter.

\subsection{Experimental setup}

The quadcopters used in our experiments are localized via sensor fusion of a motion capture system and an onboard rate gyroscope.
Experimental data from the motion capture system, voltage sensor, and current sensor are logged via radio for post-processing.
We control the quadcopters using a cascaded PID position and attitude controller shown in Fig. \ref{fig:ctrler}.
The integral action on the position and yaw helps prevent steady state errors and ensures that the vehicles are correctly vertically aligned.
Additionally, feedforward thrust for the main quadcopter to reject aerodynamic disturbances is directly added to the total thrust based on the relative location of the flying battery with respect to the main quadcopter.

\subsection{Demonstration}\label{sec:demonstration}

To demonstrate the ability and flight time benefit of our design, we conduct the following experiment:
\begin{enumerate}
	\item The main quadcopter takes off with a fully charged primary battery and hovers at a specific desired position.
	\item A fully charged flying battery is commanded to dock on the main quadcopter.
	\item Once docked, the main quadcopter switches its power source to the secondary battery and continues hovering.
	\item Once the secondary battery is completely discharged, the main quadcopter switches back to primary battery.
	\item The flying battery is commanded to undock and land on the ground. Simultaneously, another fully charged flying battery takes off.
	\item The second flying battery docks on the main quadcopter and we again switch to the secondary battery.
	\item During this period, we manually replace the discharged flying battery with a fully charged one.
	\item This process is repeated until the primary battery of the main quadcopter, only consumed during the undocking and docking process, is completely discharged.
\end{enumerate}

Fig. \ref{fig:dockTwoMQReel} shows steps (4)-(6) of the procedure.
In this demonstration, the main quadcopter hovered for a total time of \SI{57}{\minute}.
Hovering time of the main quadcopter flying alone without the dock-switch-undock-repeat process is \SI{12}{\minute}.

\subsection{Discussion}\label{sec:discussion}
The plots of input voltage and current vs. time of the main quadcopter for the duration of the entire demonstration are shown in Fig. \ref{fig:battVI}.
We see the characteristic LiPo battery discharge curve \cite{navarathinam2011characterization} several times in the voltage vs. time plot.
Each shows the complete energy consumption of one secondary battery.
The current vs. time plot shows that current input to the quadcopter increases as the voltage decreases.
This is expected because the power consumption of the quadcopter must remain approximately constant to hover continuously.

The regions in Fig. \ref{fig:battVI} highlighted in yellow show the parts of the experiment where the secondary battery did not connect to the circuit after docking.
In those regions, the main quadcopter continues using the primary battery.
We command the incorrectly docked flying battery to undock and another fully charged flying battery to dock.

\begin{figure}
	\centering
	\includegraphics[width=\columnwidth]{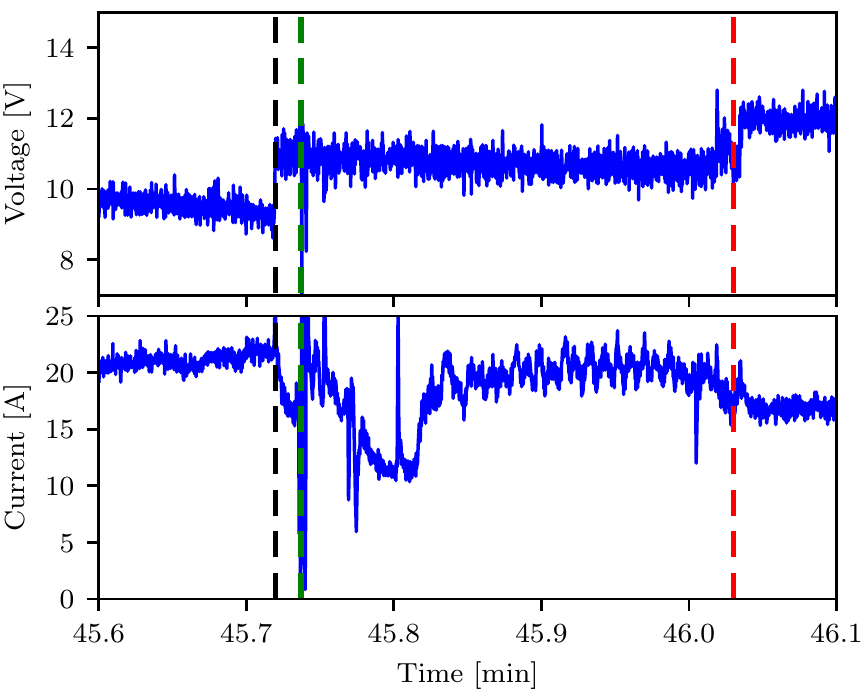}
	\caption{Zoomed version of green highlighted region in Fig. \ref{fig:battVI} showing the input voltage, current, and power of the main quadcopter during undocking and docking.
		Dotted lines mark the following events:
		(i) black: main quadcopter switches to primary battery,
		(ii) green: flying battery undocks,
		(iii) red: another flying battery docks.}
	\label{fig:battVIZoomed}
\end{figure}

A typical undocking and docking maneuver part is highlighted in green on the plot and a zoomed-in version of that region is shown in Fig. \ref{fig:battVIZoomed}.
When the main quadcopter switches back to the primary battery, the input voltage jumps to the primary battery voltage.
If the relay is closed, the main quadcopter draws power from the battery which is at a higher voltage.
The flying battery then undocks.
A current (and hence, power) surge is observed because the main quadcopter is now rejecting aerodynamic forces.
This is followed by the flying battery moving out and landing.
In this part, the main quadcopter is flying without any additional mass or aerodynamic disturbances and hence we observe a dip in power consumption.
A few seconds later, another fully charged flying battery flies on top of the main quadcopter and begins descending to dock.
We again observe an increase in power consumption because of aerodynamic force rejection.
Lastly, the flying battery docks on the main quadcopter.
We see another jump in voltage because the secondary battery is fully charged, and at a higher voltage than the primary battery.
Power consumption now settles around a value needed for hovering in the docked configuration.

The reliability of the docking contact under maneuvering, as analyzed in Section~\ref{sec:dockMech}, was tested by flying the main quadcopter powered by the flying battery in an oscillating motion with peak lateral accelerations of {\SI{12}{\meter \per \square \second}} without losing contact.
This is shown in the attached video.

From Table~\ref{table:vehicleSpec} the fraction of the main quadcopter's total mass from the battery is $\phi\approx0.23$, with the vehicle's base mass $\baseMass=\SI{630}{\gram}$.
The vehicle's flight time is approximately \SI{12}{\minute}.
Referring to Fig.~\ref{fig:battWt}, this mass fraction corresponds to a total flight time of approx. 0.47 times the optimal flight time achievable for this vehicle.
Thus, an ``optimal'' design would be capable of a maximum flight of \SI{25.6}{\minute}.
This design would however require a battery of \SI{1.26}{\kilogram} for a total vehicle mass of \SI{1.89}{\kilogram}, increasing the vehicle's overall mass by more than a factor of two.

The flying battery concept is thus able to carry the vehicle payload for a flight more than twice as long as the theoretical limit, while maintaining the total vehicle mass low, thus resulting in a safer, more useful vehicle.

\section{Conclusions} \label{sec:conclusion}

In this paper, we have introduced the concept and design of a flying battery - a small quadcopter that can carry a secondary battery for a main quadcopter, dock on it, allow it to switch its power supply from the primary battery to the secondary battery and back, and undock from it.

We designed a passive mid-air docking mechanism in the form of docking legs on the flying battery which mate with a docking platform on the main quadcopter.
We also designed a battery switching mechanism to seamlessly switch the power source of the main quadcopter mid-flight.
This was achieved using diodes to avoid backflow of current into batteries, a relay to turn the primary supply off or on, and spring loaded connectors and copper plates for electrical connection to draw power from the secondary battery.

We utilized an empirical model to provide a feedforward thrust from the main quadcopter to reject the aerodynamic disturbance forces due to the downwash of the flying battery.

Lastly, we demonstrated the ability of the system to dock, switch batteries, and undock multiple times in a single flight.
This helped the main quadcopter achieve a flight time of \SI{57}{\minute} as compared to its solo flight of \SI{12}{\minute}.
This is a 4.7-fold increase in the flight time, and a $2.2\times$ increase over the theoretical flight time limit, all while keeping the vehicle in essentially the same safety class.
This can be extremely useful in, for example, continuous monitoring activites.


An extension to this work is to redesign the onboard circuitry to make secondary batteries recharge the primary battery.
This removes the constraint of fully discharging the primary battery and, in principle, should give us unlimited flight time.
The new constraint would be the number of charging cycles that the primary battery can handle.

Another extension is exploring range extension with this concept, including the more complex challenge of docking and undocking while the main quadcopter is moving.

A third extension is to use only on-board sensing for the docking, rather than relying on an external motion capture system as done in the demonstrations in this paper.
Absolute localization may be performed via sensor fusion of INS and GPS \mbox{\cite{nemra2010robust}}.
A camera mounted on the main quadcopter and markers pasted on the flying batteries may be used for computer vision based localization \mbox{\cite{sani2017automatic, bacik2017autonomous}} as it offers high precision relative localization suitable for the sensitive docking and undocking maneuvers.

\section*{Acknowledgment}
We gratefully acknowledge financial support from NAVER LABS.
The experimental testbed at the HiPeRLab is the result of contributions of many people, a full list of which can be found at \url{hiperlab.berkeley.edu/members/}.
The authors wish to acknowledge Minos Park for assisting with the experimental validation. 

\balance
\bibliographystyle{IEEEtran}
\bibliography{main}

\end{document}